\documentclass[12pt]{amsart}
\usepackage{graphicx,amssymb,amsfonts,epsfig,amsthm,a4,amsmath,url}
\usepackage{mathrsfs}
\input amssym.def
\input amssym

\chardef\bslash=`\\ 

\makeatletter
\def\verbatim{\interlinepenalty\@M \@verbatim
   \leftskip\@totalleftmargin\advance\leftskip2pc
   \frenchspacing\@vobeyspaces \@xverbatim}
\makeatother
\newtheorem{thm}{Theorem}[section]
\newtheorem{cor}[thm]{Corollary}
\newtheorem{lem}[thm]{Lemma}
\newtheorem{prop}[thm]{Proposition}

\theoremstyle{definition}
\newtheorem{defn}{Definition}[section]

\theoremstyle{remark}
\newtheorem{rem}{Remark}[section]
\newtheorem{exmp}{Example}[section]

\numberwithin{equation}{section}

\newcommand{\begeq}{\begin {equation}}
\newcommand{\eq}{\end{equation}}
\newcommand{\bs}{\begin {split}}
\newcommand{\es}{\end{split}}
\newcommand{\bp}{\begin {prop}}
\newcommand{\ep}{\end {prop}}
\newcommand{\bt}{\begin {thm}}
\newcommand{\et}{\end {thm}}
\newcommand{\bc}{\begin {cor}}
\newcommand{\ec}{\end {cor}}
\newcommand{\bl}{\begin {lem}}
\newcommand{\el}{\end {lem}}
\newcommand{\bpf}{\begin {proof}}
\newcommand{\epf}{\end {proof}}
\newcommand{\bi}{\begin {itemize}}
\newcommand{\ei}{\end {itemize}}
\newcommand{\ben}{\begin {enumerate}}
\newcommand{\een}{\end {enumerate}}
\newcommand{\brem}{\begin {rem}}
\newcommand{\erem}{\end {rem}}
\newcommand{\bex}{\begin {exmp}}
\newcommand{\eex}{\end {exmp}}
\newcommand{\bd}{\begin {defn}}
\newcommand{\ed}{\end {defn}}

\newcommand{\la}{\langle}
\newcommand{\ra}{\rangle}

\newcommand{\ZZ}{{\mathbb Z}}

\newcommand{\CC}{{\mathbb C}}
\newcommand{\NN}{{\mathbb N}}

\newcommand{\KK}{{\mathcal K}}
\newcommand{\AK}{{\mathcal {AK}}}

\DeclareMathOperator{\supp}{supp}

\begin{document}

\title
{Krylov subspace methods in dynamical sampling}

\author{Akram Aldroubi and Ilya Krishtal}
\address{Dept. of Mathematics, Vanderbilt University, Nashville, TN 37240 \\
email: aldroubi@math.vanderbilt.edu}
\address{Department of Mathematical Sciences, Northern Illinois University,
Watson Hall 320,
DeKalb, IL 60115 \\ email: krishtal@math.niu.edu}

\thanks{The authors were supported in part by the collaborative NSF ATD grant DMS-1322099 and DMS-1322127.}

\date{\today }




\begin{abstract}
Let  $B$ be an unknown linear evolution process on $\CC^d\simeq\ell^2(\ZZ_d)$  driving an unknown initial state $x$ and producing the states $\{B^\ell x, \ell = 0,1,\ldots\}$ at different time levels.  The problem under consideration in this paper is to find as much information as possible about $B$ and $x$ from the measurements $Y=\{x(i)$, $Bx(i)$, $\dots$, $B^{\ell_i}x(i): i \in \Omega\subset \ZZ^d\}$.  If $B$ is a ``low-pass'' 
convolution operator, we show that we can  recover both $B$ and $x$, almost surely, as long as we  double the amount of temporal samples needed in \cite{ADK13} to recover the signal propagated by a known operator $B$. For a general operator $B$, we can recover parts or even all of its spectrum from $Y$. As a special case of our method, we derive  the centuries old Prony's method \cite{BDVMC08, P795, PP13} which recovers a vector with an $s$-sparse Fourier transform from $2s$ of  its consecutive components. 
\end{abstract}
\maketitle

\section{Introduction.}
 Sampling of physical processes is done by sensors or measurement devices that are placed at various locations and can be activated at different times. Dynamical sampling involves studying the time-space patterns formed by  the locations of the measurement devices 
 and the times of their activation \cite {ADK13, ADK14,ACMT14,LV09, RMG12, GRUV14}.  One of the goals in dynamical sampling is to identify the patterns that allow one to deduce the desired information about the evolution process.
 In \cite{ADK13, ADK14}
we considered the problem of spatiotemporal sampling in which an initial state of an evolution process  is to be recovered from a set of samples at different time levels. There it is assumed that the 
evolution process is driven by a known well-behaved filter arising from a well-studied physical process such as diffusion. 
In 
\cite{AT14, ACMT14, D14}, the problem of signal recovery was studied for more general processes but the
operator governing the evolution was still assumed to be known. 
In this paper, we shift the emphasis to recovering the spectrum of the evolution operator, which is no longer assumed to be known. In the case of \cite{ADK13}, the operator is defined by a well-behaved filter and, thus,
is completely determined by its spectrum.
 
 In \cite{ADK13, ADK14} we have shown that it is possible to trade off spatial samples for time samples  at essentially a one-to-one ratio without any loss of information. In this paper
 we show that if the number of time samples in an invariant process is doubled, we can still solve the problem even if the filter propagating the signal is not known {\it a priori}. For more general processes, we may not be able to recover the evolution operator completely, but we are still able to recover part or all of its spectrum.

Dynamical sampling setup is in many ways similar to that of the Slepian-Wolf distributed source coding problem \cite{SW73} and the distributed sampling problem in \cite{HRLV10}. Our setting, however,  is fundamentally different from the above in the nature of the processes we study. Distributed sampling problem typically deals with two signals correlated by a transmission channel. We, on the other hand, can observe an evolution process   
at several instances and
over longer periods of time. In the invariant case, we gain access to a number of signals (observations) correlated via the same filter. We then employ a reincarnation of a well-known Prony's method \cite{BDVMC08, P795, PP13} that uses these observations first for the almost sure recovery of the correlating filter and next for the recovery of the initial state of the process as in  \cite{ADK13}. Intuitively, one can think of recovering of the shape of a wave by observing its amplitude at a single location over a long period of time as opposed to acquiring all of the amplitudes at once. For more general processes, we develop more sophisticated Krylov subspace methods
to recover (a part of) the spectrum of an evolution operator. 

Let us introduce some of the relevant notation and describe our problem in more detail.
In this paper we limit our attention to the finite dimensional case and expect to address the infinite dimensional
analog elsewhere. 

As in \cite{ADK13},
the \emph{signal} $x$ here is  represented by a vector in
 $\ell^2(\ZZ_d)\simeq \CC^d$, where $\ZZ_d$ is the cyclic group of order $d\in\NN$. The \emph{evolution
 operator} $B\in B(\ell^2)$ is represented by a $d\times d$ matrix. 
We call the evolution
 process \emph{invariant} if $B$ is a self-adjoint circular matrix. 
 
 The \emph{sampling operator} or \emph{sampler} $A\in B(\ell^2)$ is another $d\times d$ matrix. An \emph{ideal sampler}  $A=S_\Omega$ is the diagonal projection
 \[
 S_{\Omega}x = \sum_{j\in\Omega} \la x, e_j\ra e_j,\quad x\in\CC^d,
 \]
 where $\{e_j, j=1,\ldots, d\}$ is the standard orthonormal basis in $\CC^d$. If $m$ divides $d$
 and $\Omega = \Omega_m = \{0, m, 2m, \dots\}$, we shall write $S_m$ instead of $S_{\Omega_m}$
 and call it a \emph{uniform} (ideal) sampler.
 
 By the \emph{dynamical samples} of a signal $x\in\CC^d$ we mean the collection of vectors
 \begeq\label{observ}
y_\ell = AB^{\ell}x,\ \ell = 0, 1, \ldots 
\eq

The general goal of the research in this paper is to recover as much information about the evolution operator $B$
as possible from the dynamical samples $y_\ell$. Given sufficiently many dynamical samples, we will provide a method to recover all the eigenvalues of $B$ that can possibly be recovered from these samples. Moreover, in the case of an invariant process and a uniform sampler, we describe an algorithm to recover the operator $B$ 
completely from $2m$ dynamical samples for almost every (unknown) signal $x\in\CC^d$. The vector $x$ can then be recovered as in \cite{ADK13}.

The paper is organized as follows. In Section \ref {sec_gen}, we introduce  the theory that connects the minimal (annihilating) polynomial of a matrix $B$ to several other types of annihilating polynomials such as the A-altered minimal polynomial of $B$ and annihilators of certain altered Krylov subspaces. This connection together with an algorithm for finding the annihilator of an $A$-altered Krylov subspace will allow us, in Section \ref {SRDM}, to find  (a part of) the spectrum  of the operator $B$ from measurements obtained by sampling $B^\ell x$, $\ell=0,\dots,\ell_i$ at locations  $i\in \Omega \subset \ZZ_d$, where $x$ is an unknown vector in $\CC^d$.   Section \ref {RIDS} is devoted to the special case when the operator $A$ is a convolution operator, as is common in applications. 


\section{Krylov subspaces and annihilating polynomials}\label{sec_gen}

Around 1930 the engineer and scientist Alexei Nikolaevich Krylov used Krylov subspaces
to compute the coefficients of a minimal polynomial, the roots of which are eigenvalues
of a matrix \cite{K31}. In this section we show that Krylov's idea can still be used in the case
of altered (or preconditioned) Krylov subspaces that we introduce. 
The definitions and the general theory developed here will be used in subsequent sections. 
As before, $A$ and $B$ are $d\times d$ complex matrices.

\bd
A \emph{Krylov subspace} of order $r$ generated by the matrix $B$ and a vector $x\in\CC^d$ is
\[\mathcal K_r(B,x) = {\rm span} \{x, Bx, \ldots, B^{r-1}x\}.\]
\ed

\bd
An \emph{altered Krylov subspace} of order $r$ generated by  matrices $A, B$ and a vector $x\in\CC^d$ is
\[{\mathcal {AK}}_r(A; B,x) = A\mathcal K_r(B,x) = {\rm span} \{Ax, ABx, \ldots, AB^{r-1}x\}.\]
\ed

We consider the following minimal annihilating polynomials with regard to Krylov and altered Krylov subspaces.

\bd\label{defminpoly} ${ } $ 
\begin{enumerate}
\item The {\em minimal  polynomial of} $B$, denoted by $p^B$, is the monic polynomial  of smallest degree among all the polynomials $p$ such that $p(B)=0$. We will denote the degree of $p^B$ by $r^B$. 

\item The {\em $A$-altered minimal polynomial of $B$}, denoted by $p^B_A$,  is the  monic polynomial of smallest degree among all the polynomials $p$ such that $Ap(B) = 0$. We  let $r^B_A$ denote the degree of $p^B_A$. 

\item The {\em $(A,B,x)$-annihilator}, denoted by  $p^B_{A,x}$, is the monic polynomial of smallest degree among all the polynomials $p$ such that $Ap(B)\KK_{r^B_{A}}(B,x) =\{0\}$,
$x\in\CC^d$. We let $r^B_{A,x}= \deg p^B_{A,x}$.
\end{enumerate}
\ed

Observe that the degrees of the minimal polynomials correspond to the orders of the respective maximal Krylov and altered Krylov subspaces. We also note that the above definition implies that if $x\in \CC^d$ is such that $\AK_{r^B_A}(B,x) = \{0\}$ then the $(A,B,x)$-annihilator is \emph{trivial}, i.e.~$p^B_{A,x} \equiv 1$.

\bp
 The polynomial $p^B_A$ always divides $p^B$.
\ep
\bpf
Since $Ap^B(B) = 0$, we have $r^B\ge r^B_A$. Then $p^B=p^B_Ag+h$ where $h$ is a polynomial of degree less than $r^B_A$, and $0=Ap^B(B)=Ap^B_A(B)g(B)+Ah(B)$ implies $Ah(B)=0$. But then $h = 0$ due to the minimality
of $p^B_A$.
\epf
\bc
\label {rootspba}
The roots of $p^B_A$ are eigenvalues of $B$.
\ec

The proposition above can be restated in terms of ideals generated by minimal polynomials, so that we have the following  inclusion $\langle p^B\rangle \subseteq \langle p^B_A\rangle $. The next  lemma is key to the rest of this section and states that the set of all the polynomials $p$ such that $Ap(B)\KK_{r^B_{A}}(B,x) =\{0\}$  is an ideal generated by  $p^B_{A,x}$.  Note that the polynomials $p$ such that $Ap(B)\KK_{L}(B,x) =\{0\}$ for $L<r^B_{A}$ do not necessarily form an ideal. 
\bl\label{mainlemma}
Assume that a polynomial $q$ satisfies $Aq(B)\KK_{r^B_A}(B,x) =\{0\}$. Then 
for any $\ell\in\NN$ we have $Aq(B)\KK_{\ell}(B,x) =\{0\}$. Equivalently,
for any polynomial $s$ we have
$Aq(B)s(B)x = 0$.
\el

\bpf
Let $s$ be a polynomial. If $\deg s < r^B_A$, then the conclusion is contained in the assumption. Otherwise, we represent $s$ as $s = p^B_Ag+h$ where $g$ and $h$ are polynomials such that $\deg h < \deg p^B_A = r^B_A$. 
Then $Aq(B)s(B)x = Aq(B)(p^B_A(B)g(B)+h(B))x = Ap^B_A(B)q(B)g(B)x + Aq(B)h(B)x =  0$.
\epf
As a consequence we get the following inclusion of ideals $ \langle p^B_A\rangle  \subseteq \langle p^B_{A,x}\rangle $. 
\bp
\label {abxann}
The $(A, B, x)$-annihilator $p^B_{A,x}$ divides $p^B_A$, and, hence, the roots
of $p^B_{A,x}$ are eigenvalues of $B$.
\ep

\bpf
We only need to prove something if $r^B_{A,x} > 0$.
Directly from Definition \ref{defminpoly} we have
$r^B_{A,x} \leq r^B_A$. Hence, we can write $p^B_A = p^B_{A,x}g+h$ where $g$ and $h$ are polynomials and   $\deg h <  r^B_{A,x}$.
Using Lemma \ref{mainlemma} we have
\[
\{0\} = Ap^B_A(B)\KK_{r^B_A}(B,x) = A(p^B_{A,x}(B)g(B)+h(B))\KK_{r^B_A}(B,x) = Ah(B)\KK_{r^B_A}(B,x)
\]
and the minimality of $p^B_{A,x}$ implies $h = 0$.
\epf
The next proposition is important for the development of the rest of this paper, and it states that the ideals   $\langle p^B_A\rangle$ and  $\langle p^B_{A,x}\rangle$ are equal for almost all $x$.
\bp \label {MIC}
The set $\Upsilon = \{x \in \CC^d:\ p^B_{A,x} \neq p^B_A\}$ has Lebesgue measure $0$.
\ep

\bpf 
Let $q$ be a monic polynomial of degree less than $r^B_A$ that divides $p^B_A$. There are finitely many such polynomials.  The set  $\Upsilon_q = \{x\in \CC^d:\  p^B_{A,x}=q\}$ is a  subspace of $\CC^d$ and
$\Upsilon_q\neq \CC^d$ since $p^B_A$ is minimal. Hence, $\Upsilon_q$ has measure $0$. So does the set $\Upsilon = \bigcup_q \Upsilon_q$ which is a finite union of null sets.
\epf

The above results show that the spectral identification in dynamical sampling hinges on the ability to compute the
$(A, B, x)$-annihilator. The following result identifies the number of dynamical samples that is sufficient for the computation.

\bp\label{aver}
Assume $r^B_{A,x} > 0$. Then the $(A, B, x)$-annihilator can be recovered from $2r^B_A$ dynamical samples 
$y_\ell = AB^{\ell}x$, $\ell = 0, 1, \ldots, 2r^B_A-1$. 
\ep

\bpf
For $r_x \in\NN$, consider the system
\[
AB^{r_x}B^k x+\sum_{\ell=0}^{r_x-1}\alpha_\ell AB^{\ell}B^{k}x = 0, \quad k=0,\ldots, r^B_A-1,
\]
of linear equations in $\alpha = (\alpha_0, \ldots \alpha_{r_x-1})$. Clearly, the only $r_x$ for which the above system has a unique solution equals $r^B_{A,x}$ and $\alpha$ is the vector of the coefficients of the
$(A, B, x)$-annihilator. Since $r^B_{A,x} \le r^B_A$, the samples 
$y_\ell = AB^{\ell}x$, $\ell = 0, 1, \ldots, 2r^B_A-1$,
provide enough information to explicitly write the above system when $r_x = r^B_{A,x}$.
\epf

\bex
\label {prony1}
In this example we let the evolution operator $B$ be the circular shift, $(Bx)(n) =x(n+1)$, $x\in \ell^2(\ZZ_d)$, and the sampler $A$ be equal to $S_{\{j\}}$ for some $j\in\ZZ_d$. 
Observe that $\sigma(B) = \{e^{2\pi i \frac nd}: n\in\ZZ_d\}$.
We shall assume that the signal $x$ is such that its Fourier transform $\hat x$ is $s$-sparse, i.e.~the cardinality of $\supp \hat x$ is $s<\frac{d}2$. Let $V$ be the subspace $V=\{y\in \CC^d: \supp \hat y\subseteq\supp\hat x\}$. Since $V$ is invariant under $B$, we see that $r^B_{S_{\{j\}},x} = s$ and the system of $2s$ equations as in the proof of Proposition \ref{aver} gives
the coefficients of the annihilator $p^B_{S_{\{j\}},x}$ as its solution. The roots of the annihilator are the $d$-th roots of unity and, moreover, $e^{2\pi i \frac nd}$ is a root if and only if $n\in\supp\hat x$. These roots constitute the spectrum $\sigma(B\vert_V)$ of the restriction $B\vert_V$ of $B$ to $V$.
 Thus, for this case, the spectral recovery in dynamical sampling is equivalent to the well-known Prony's method {\cite{BDVMC08, P795}} for the recovery of a vector with an $s$-sparse Fourier transform from $2s$ of its consecutive samples. In fact, working out the details, one sees that our algorithm is the same as that of Prony (see Example \ref{prony2} below). A closely related point of view on the Prony's method and its generalizations is presented in \cite{PP13}.
\eex
\brem
Let us emphasize  that in the above example, the vector  $x$ belongs to the set $\Upsilon$ defined in Proposition \ref {MIC}. 
\erem

\brem
In numerical linear algebra the modern Krylov
subspace methods are typically based on the Lanczos or Arnoldi processes \cite[and references therein]{W07}. In our future research we plan to investigate 
if any of these methods can be modified to solve the problem of spectral identification in  dynamical sampling.
\erem

\section{Spectral recovery for diagonalizable matrices}
\label {SRDM}
 
To simplify the exposition, in the remainder of this paper we consider only diagonalizable evolution operators.
Thus, we write $B = UDU^{-1}$, where $D$ is a diagonal matrix and $U$ is an invertible matrix. 
The columns of the matrix $U^*$ will be denoted by $u_i= U^*e_i$, $i=1,\ldots,d$, where, as usually, $\{e_j, j=1,\ldots, d\}$ is the standard orthonormal basis in $\CC^d$.
 We let  $\sigma(B) = \sigma(D) = \{\lambda_1,\ldots,\lambda_n\}$ be the spectrum of the matrix $B$, which we seek to recover from the dynamical samples. We write the spectral decomposition of the matrix $D$ as
 $D = \sum_{j=1}^n \lambda_j P_j$.
 
\bd
Let $\Omega$ be a subset of $\{1,\ldots, d\}$.
An eigenvalue $\lambda_j\in \sigma(B)$ is called \emph{$\Omega$-observable} if $S_\Omega U P_j\ne 0$.
The set of all $\Omega$-observable eigenvalues will be denoted by $\sigma_\Omega(B)$.
\ed

In this section, we show that, for almost every vector $x\in\CC^d$,  an eigenvalue $\lambda_j$ can be recovered
from the dynamical samples 
if and only if it is $\Omega$-observable.

 
We begin by considering an important special case 
which occurs when the sampler $A = S_{\{i\}}$ is the rank $1$ orthogonal projection onto the $i$-th basis element. 
Recall that we have defined $u_i = U^*e_i$.

\bl
\label {DualPi}
The  $(I, D^*, u_i)$-annihilator $p^{D^*}_{I,u_i}$
coincides with the polynomial $\tilde {p} ^B_{S_{\{i\}}}$ whose coefficients are complex conjugates
of the coefficients of the $S_{\{i\}}$-altered minimal polynomial $p^B_{S_{\{i\}}}$.
\el

\bpf
The claim follows from the equalities
\begeq\label{keyadjoint}
\la S_{\{i\}}B^k x, e_i\ra = \la B^k x, e_i\ra = \la UD^kU^{-1}x, e_i \ra = \la U^{-1}x, (D^*)^ku_i\ra,
\eq 
$k\in\NN$, $x\in\CC^d$, and the fact that $\la S_{\{i\}}B^k x, e_j\ra =0$ for all $j\ne i$. 
\epf

\bl
\label {eqspec} Let 
$R(p^{D^*}_{I,u_i})\subset \CC$ be the set of all roots of the $(I, D^*, u_i)$-annihilator $p^{D^*}_{I,u_i}$. Then 
$\sigma_{\{i\}}(B)=\overline{R(p^{D^*}_{I,u_i})}$, where the bar denotes complex conjugation.
\el
\bpf
To simplify the notation, in this proof we will write $q$ instead of $p^{D^*}_{I,u_i}$.
First, observe that for $j=1,\ldots,n$ we have $\lambda_j\in \sigma_{\{i\}}(B)$ if and only if $P_jU^*S_{\{i\}}\ne 0$, and the latter inequality is equivalent to $P_ju_i=P_jU^*e_i=P_jU^*S_{\{i\}}e_i\ne0$.  

Secondly, observe that for  any polynomial $p$ we have 
\[p(D^*)u_i=\sum_{j=1}^np(\overline{\lambda_j})P_ju_i,\]
and the collection of vectors $\{P_ju_i: P_ju_i\neq 0\}$ is linearly independent.
In particular, if $p=q$, then  $0=\sum_jq(\overline{\lambda_j})P_ju_i=0$ implies $q(\overline{\lambda_j})=0$ for all $\lambda_j \in \sigma_{\{i\}}(B)$. Hence, $\sigma_{\{i\}}(B)\subseteq \overline{R(q)}$.

Finally, the polynomial $h$ defined by
\[h(\lambda) =\prod_{j:  P_ju_i\neq 0} (\lambda-\overline{\lambda_j})
= \prod_{j\in \sigma_{\{i\}}(B)} (\lambda-\overline{\lambda_j})\]
satisfies $h(D^*)u_i = 0$. Hence, since $\sigma_{\{i\}}(B)\subseteq \overline{R(q)}$ and $q$ is minimal,
we must have $h=q$.
\epf
As a corollary of Lemmas \ref {DualPi} and \ref {eqspec}  we immediately get 
\bc 
\label {corcard} We have
$r^B_{S_{\{i\}}} = r^{D^*}_{I,u_i}=| \sigma_{\{i\}}(B)|$, where $| \sigma_{\{i\}}(B)|$ is the cardinality
of the set.
\ec

The following result is an immediate extension of the above observations to all ideal samplers.

\bp
Let $p^{D^*}_{\Omega}$ denote the least common multiple of the $(I, D^*, u_i)$-annihilators $p^{D^*}_{I,u_i}$,
$i\in\Omega$. Then the coefficients of $p^{D^*}_{\Omega}$ coincide with the complex conjugates of the coefficients 
of the $S_{\Omega}$-altered minimal polynomial $p^B_{S_{\Omega}}$.
\ep
\bpf
This follows directly from the fact that $S_\Omega =\sum\limits_{i\in \Omega}S_{\{i\}}$ and the minimality of $p^B_{S_{\Omega}}$.
\epf

\bc \label {EqCard} We have
$r^B_{S_{\Omega}} = | \sigma_{\Omega}(B)|$.
\ec

Combining Proposition \ref{MIC} and corollary \ref {corcard}   with Proposition \ref{aver} we get
\bt\label {finthm21}
For almost every $x\in\CC^d$ the set $\sigma_{\{i\}}(B)$ 
can be recovered from the measurements $\{(B^kx)(i): k=0,\ldots, 2r_i-1\}$, 
where $r_i = |\sigma_{\{i\}}(B)|$.
Consequently, for any  $\Omega\subseteq \ZZ_d$, 
the set  $\sigma_\Omega(B)$ can be recovered  from the dynamical samples $\{(B^kx)(i)$: $ i \in \Omega$, $k=0,\ldots, 2r_i-1\}$ for almost every $x\in\CC^d$. 
\et
\brem
From the proof of Lemma \ref {eqspec} we know that  $\lambda_j\in \sigma(B)$ is  $\Omega$-observable if and only if there exists $i\in \Omega$ such that $P_ju_i\ne 0$, where $u_i= U^*e_i$. If $\Omega$ is a set that allows the reconstruction of any vector $x\in\CC^d$ from samples $\{(B^kx)(i)$: $ i \in \Omega$, $k=0,\ldots, 2r_i-1\}$, then, from the results in \cite {ACMT14}, $\{P_ju_i: i \in \Omega\}$ is a frame for the range of $P_j$. Thus, it follows that if $\Omega$ is a set that allows the reconstruction of any  $x\in\CC^d$ from samples $\{(B^kx)(i)$: $ i \in \Omega$, $k=0,\ldots, 2r_i-1\}$, then $\sigma_\Omega(B)=\sigma {(B)}$.
\erem

We conclude the section with an estimate on the number of time samples needed for the recovery of the spectrum under an additional  condition.

\bt
\label {finthm3}
Let $\{u_i: i\in \Omega \}$ be the column vectors of the matrix $U$ corresponding to $\Omega$ and let $L$ be a fixed integer. Assume that 
$\{(D^*)^Lu_i: i \in \Omega\} \subset span 
\{u_i,D^*u_i,\dots, (D^*)^{L-1}u_i: i\in\Omega\}$.
Then   the set $\sigma_\Omega(B)$ can be recovered  from  $\{(B^k)x(i): i  \in \Omega, k=0,\ldots, (|\Omega|+1)L-1\}$
for almost every $x\in\CC^d$.
\et

\bpf
The assumption of the theorem implies that there exist numbers $\alpha_{\ell j}(i)$, $i,j\in\Omega$, $\ell=0, \ldots, L-1$, such that
\[(D^*)^Lu_i = \sum_{j\in\Omega}\sum_{\ell=0}^{L-1} \alpha_{\ell j}(i) (D^*)^{\ell}u_j,\quad i\in\Omega.\]
From \eqref{keyadjoint} it follows that for almost every $x\in \CC^d$ we have
\begeq\label{keyrecover}
(B^{L+k}x)(i) =  \sum_{j\in\Omega}\sum_{\ell=0}^{L-1} \overline{\alpha_{\ell j}(i)} (B^{\ell+k}x)(j), \quad i\in\Omega,\ k\in\ZZ.
\eq
Observe that for $k=0,\ldots, |\Omega|L-1$ and $i\in\Omega$ the equations in \eqref{keyrecover} form a square system of linear equations which has a solution by the assumption of the theorem. Computing the coefficients 
$\alpha_{\ell j}(i)$, $i,j\in\Omega$, $\ell=0, \ldots, L-1$, from these square systems and plugging them back into
 \eqref{keyrecover} we obtain $(B^{k}x)(i)$ for any $k\in\ZZ$ and $i\in\Omega$. It remains to invoke Theorem \ref{finthm21}.
\epf
\brem
Note that in many cases we have that $|\Omega|L$ is approximately $d$. Thus, a typical number of measurements needed to obtain the spectrum via Theorem \ref {finthm3} is approximately $(|\Omega|+1)d$.
\erem
\brem
Theorems \ref {finthm21} and \ref {finthm3} are also valid for general matrices $B$ with some minor modifications. In particular, Theorem \ref {finthm3} remains valid if the Jordan decomposition $B=UJU^{-1}$ is used, and the matrix $D$ is  replaced by the matrix $J$ in its statement.  

\erem
\section{Recovery of the evolution operator in the case of a regular invariant dynamical sampling problem.}\label{RIDS}

An important question for spectral recovery is determining the actual number of dynamical samples 
needed to find the spectrum of $B$. Proposition \ref{aver} provides essentially the best answer one could have in the general case.
From the algorithmic point of view, however, this answer is not very satisfactory. Indeed, to  determine the number $r^B_A$ used in  Proposition  \ref{aver},  one needs to have
some prior knowledge about $B$.
In this section we address the case of an invariant evolution process with a uniform ideal sampler. This turns out to be enough prior information to get a good upper bound on the
sufficient number of dynamical samples.


Recall from the introduction,  that in the case under consideration the sampler $A$ is given by
\begeq\label{Sm}
A = S_m: \ell^2(\ZZ_d)\to  \ell^2(\ZZ_d),\quad  (S_m x)(n) = \delta_{(n\!\!\!\!\mod m), 0}x(n), 
\eq
where  $m$ is an odd integer that divides an odd integer $d$ (oddness is assumed for the sake of computational simplicity).
The evolution operator $B$ in this case is a convolution operator:
$Bx = a*x$, $a\in \ell^2(\ZZ_d)\simeq \CC^d$. Since the matrix of $B$ is circular, it is diagonalized by 
the 
$d$-dimensional discrete Fourier transform (DFT)  $ {\mathbf F}_d$ defined by
$$\hat x(k) = ({\mathbf F}_d x)(k) = 
\sum\limits_{\ell=0}^{d-1} x(\ell)e^{-\frac{2\pi ik\ell}d }, \quad x\in \CC^d,\ k=1,\dots, d.$$
Thus, we have
\begeq\label{diagfour}
B = {\mathbf F}_d^{-1} D {\mathbf F}_d = \frac1d{\mathbf F}_d ^*D {\mathbf F}_d,
\eq
where $D$ is the diagonal matrix defined by $\hat a$ -- the DFT of the filter $a$.

In this section we show that for almost every $x\in\CC^d$ the knowledge of  $2m$ dynamical samples 
\begeq\label{observ11}
y_\ell = S_mB^{\ell}x,\ \ell = 0, \ldots, 2m-1, 
\eq
is sufficient to reconstruct the spectrum of $B$, that is the set of entries of $\hat a$.


Observe that knowing the order of entries in $\hat a$ would then allow one to recover the operator $B$ completely via \eqref{diagfour}. In particular, this can be done if one assumes that
$\hat a$ is real, symmetric, and decreasing on $\{0,1,\ldots, \frac{d-1}2\}$, which is a natural assumption for diffusion-type processes. Once the
filter $a$ is known, one can use the results in \cite{ADK13} to recover $x$.

\bt\label{finthm1}
Assume  that $B$ is an unknown convolution operator  on $ \ell^2(\ZZ_d)$. Then, for almost all $x\in\CC^d$, the spectrum of $B$ can be recovered from the dynamical samples $y_\ell$, $\ell = 0, \ldots, 2m-1$, defined in \eqref{observ11}.
\et
 
\bpf 
We use the DFT and the Poisson summation formula to rewrite \eqref{observ11} in the following way:
\[
\hat y_\ell = {\mathbf F}_d S_m {\mathbf F}_d^{-1}{\mathbf F}_d B^\ell {\mathbf F}_d^{-1} \hat x =
\sum_{j=1}^J E_j D^\ell \hat x,
\]
where each $E_j$, $j=1,\ldots, J$, is a rank-1 projection given by
\begeq\label{PoissonProj}
(E_j z)(k) =  \left\{\begin{array}{rl}
\frac{1}m\sum\limits_{i=0}^{m-1}{z}(k + Ji),& k = j\mbox{ mod }J; \\
0,& \mbox{otherwise};
\end{array}
\right.
\ z\in\CC^d,\ k=1,\ldots, d.
\eq
Observe that $E_jE_k = \delta_{jk}E_k$, where $\delta_{jk}$ is the usual Kronecker delta. 
%
For $j= 1, \ldots, J$, let $\Omega(j) = \{k\in \ZZ_d: k = j\!\!\mod J\}$.
Since the $k$-th row of $E_j$ is zero for any $k\notin \Omega(j)$ and $D$ is diagonal,
the polynomials
\[
p_j(\lambda) = \prod_{k\in\Omega(j)} (\lambda - \hat a(k))
\]
satisfy $E_j p_j(D) = 0$,  $j=1,\ldots, J$. Moreover, the set $R(j)$ of all roots  of the minimal polynomial $p^D_{E_j}$ coincides with the set of 
all roots of the polynomial $p_j$ ({\it cf.}~the proof of Lemma \ref{eqspec}).
It immediately follows that  
\begeq\label{equalroots}\sigma(B) = \bigcup_{j=1}^{J} R(j)\eq  and $r^{D}_{E_j} \le m$ for each $j=1,\ldots,J$. Hence, we can apply Proposition \ref{aver} to recover the 
$(E_j, D, \hat x)$-annihilators $p^D_{E_j,\hat x}$, $j=1,\ldots, J$. Due to Proposition \ref{MIC},   these annihilators
will coincide with the minimal polynomials $p^D_{E_j}$ for almost every $x\in \CC^d$ and \eqref{equalroots}
means that the proof is complete.
%
\epf
 
\subsection {Algorithm} The proof of Proposition \ref{aver} essentially provides an algorithm for the recovery of the spectrum $\sigma(B)$. 
 For the case in Theorem \ref{finthm1}, the algorithm in Proposition \ref{aver} is unnecessarily complicated and can be simplified as follows.

Our goal is to find for each $j=1, \ldots, J$ the set of roots $R(j)$ of the $(E_j,D,\hat x)$-annihilator $p^D_{E_j, \hat x}$.
 According to the proof of Proposition \ref{aver}, 
the coefficients $\alpha(j) = (\alpha_0(j),\ldots, \alpha_{r_j-1}(j))$ of the polynomial $p^D_{E_j,\hat x}$ satisfy the system of linear equations
\begeq \label {oneq}
E_jD^{r_j}D^k \hat x+\sum_{\ell=0}^{r_j-1}\alpha_\ell (j) E_jD^{\ell}D^{k}\hat x = 0, \quad k=0,\ldots, r^D_{E_j}-1,
\eq
where $r_j$ is the degree of $p^D_{E_j, \hat x}$, i.e.~the minimal integer for which the system has a (unique) solution.

Since $r^D_{E_j}\le m$, and in view of Lemma \ref {mainlemma},  we can let $k$ vary from $0$ to $m-1$ without altering the solution of the system \eqref{oneq}.
Since $E_j$ has rank one, for each fixed $k$, the system of $d=mJ$ equations can be replaced by a single equation
\[\sum\limits_{i=0}^{m-1}\hat a^{k+r_j}(j+iJ)\hat x(j+iJ) +\sum\limits_{\ell=0}^{r_j-1}\alpha_\ell(j)\sum\limits_{i=0}^{m-1}\hat a^{k+\ell}(j+iJ)\hat x(j+iJ)=0.\]
Recalling that $\hat y_\ell(j) = \frac1m \sum\limits_{i=0}^{m-1}\hat a^{\ell}(j+iJ)\hat x(j+iJ)$, $j\in\ZZ_d$, $\ell = 0,1,\ldots$,
we deduce that the system \eqref{oneq} is equivalent to
%
\begeq \label {Fouriereq}
\hat y_{k+r_j}(j) + \sum_{\ell=0}^{r_j-1}\alpha_\ell (j) \hat y_{k+\ell}(j) =0,  \quad k=0,\ldots, m-1.
\eq
For each $j$, this last system of equations 
 can be set from the dynamical samples that are available to us by assumption. 
Thus, 
 we have the following crude algorithm for the recovery of $\sigma(B)$.

\medskip
 
 \emph{Step I}. For each $j = 1, \ldots, J$

\ben
 \item find the minimal integer $r_j$ for which the system \eqref {Fouriereq}
 has a solution $\alpha(j)$ and find that solution; 
 
\item let $p_j(\lambda) = \lambda^{r_j}+\sum_{\ell=0}^{r_j-1}\alpha_\ell (j)\lambda^\ell$ and find the set $R(j)$ of all roots of $p_j$.

\een

\emph{Step II}. Recover the spectrum $\sigma(B)$ from \eqref{equalroots}.
 
 \bex
 \label {prony2}
 Prony's method for finding vectors with sparse Fourier transform includes solving a system of linear equations that is a special case of the one in our algorithm. As in Example \ref {prony1}, we let  $(Bx)(n) =x(n+1)$ and $x\in \ell^2(\ZZ_d)$ be $s$ sparse.  We also let the subsampling factor $m$ be equal to $d$. Then the system of equations \eqref {Fouriereq} with $r_j=s$ will coincide with the system of equations for finding the $\supp \hat x$ in Prony's method.
 \eex
 
 \brem
 In \cite{ADK13} we noted that the filter $a$ is typically such that $\hat a$ is real, symmetric, and decreasing for $\ell = 0,1,\ldots \frac{d-1}2$.
In this case, it easily follows that $r_j = r^D_{E_j} = m$ for $j = 1,\ldots, J-1$, and $r_J = r^D_{E_J} = \frac{m+1}2$ for almost every $x\in\CC^d$. Moreover, the monotonicity condition allows us to properly order the roots in $\sigma(B)$ and completely recover $\hat a$ and, hence, the evolution matrix $B$. We also observe that in this case, the imaginary part of the system \eqref {Fouriereq} can be ignored and the real part of the system is given by a self-adjoint Hankel matrix. This allows one to use special methods to solve it.
  \erem
\medskip

\noindent
{\bf Acknowledgements}.
We would like to thank Nathan~Krislock for bringing Krylov subspaces to our attention. We thank Sui~Tang and Armenak~Petrosyan for insightful comments. We also thank ICERM and the organizers of the Research cluster ``Computational Challenges in Sparse and Redundant Representations'' for a wonderful opportunity to be there. Special thanks are reserved for
 S.~J.~Rose and his inspirational generosity in helping us to create this work. 

\begin{figure}\label{Rospic}
  \begin{center}
  \label {Fig1}
    \includegraphics[width = 2.4in, height =
      1.5in]{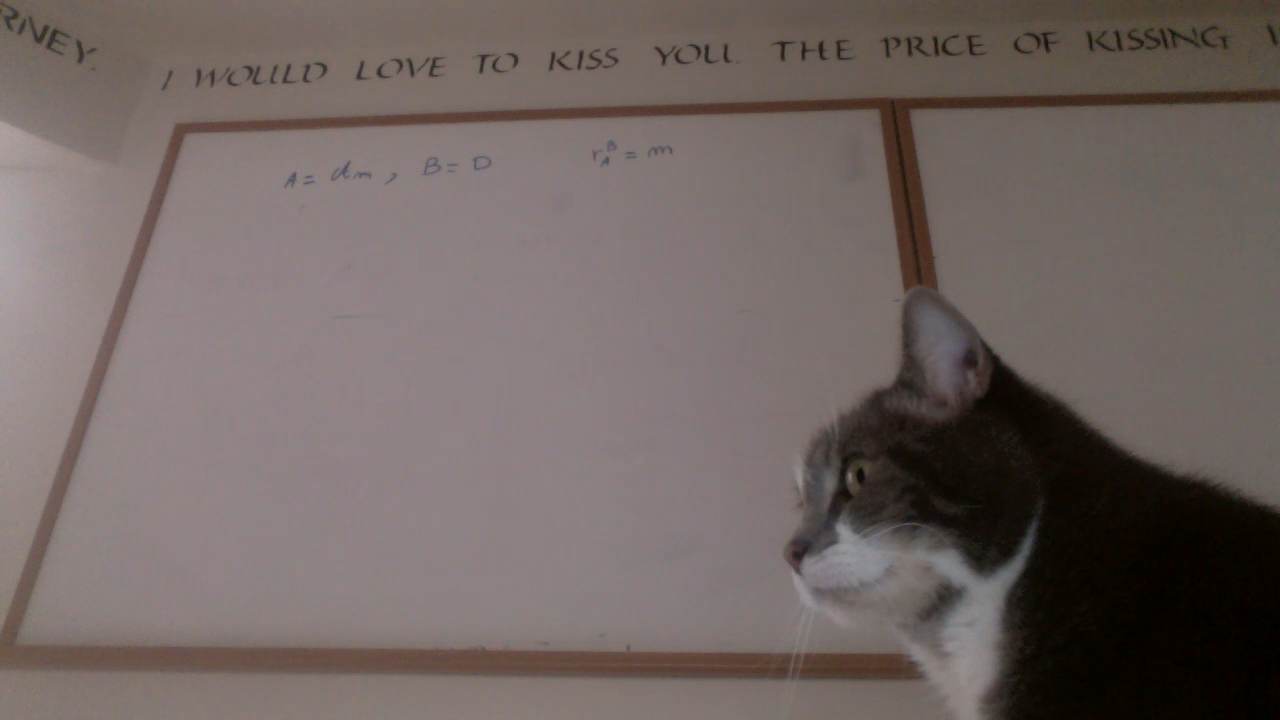}
     \ 
      \includegraphics[width = 2.4in, height =
      1.5in]{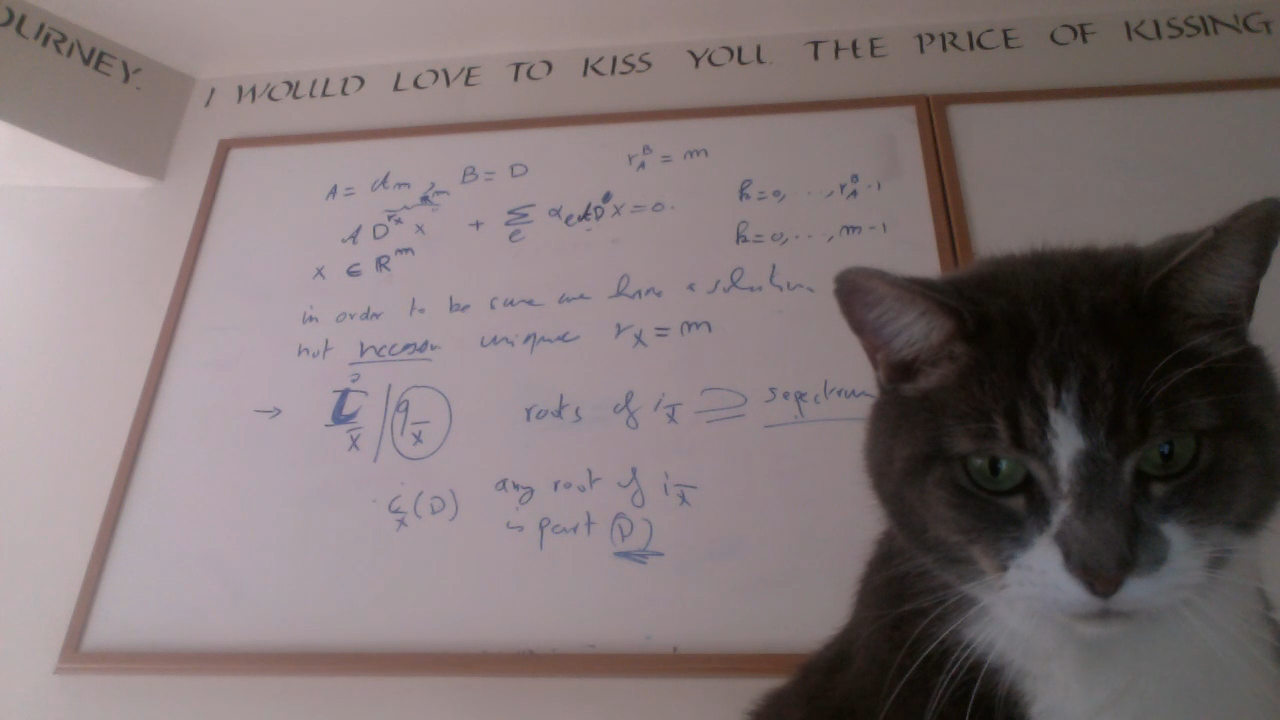}
\caption{S.~J.~Rose at work.} 
  \end{center}
\end{figure}

\bibliographystyle{siam}
\bibliography{../refs}

\end {document}